\documentclass[usegraphicx]{mn2e}
\usepackage{times}

\newif\ifAMStwofonts

\newcommand{\gtorder}{\mathrel{\raise.3ex\hbox{$>$}\mkern-14mu
                \lower0.6ex\hbox{$\sim$}}}
\newcommand{\ltorder}{\mathrel{\raise.3ex\hbox{$<$}\mkern-14mu
                \lower0.6ex\hbox{$\sim$}}}

\newcommand{\oi}{O\,{\sc i}}
\newcommand{\oii}{O\,{\sc ii}}
\newcommand{\oiii}{O\,{\sc iii}}

\newcommand{\ov}{O\,{\sc v}}
\newcommand{\ovii}{O\,{\sc vii}}
\newcommand{\oviii}{O\,{\sc viii}} 
 
\newcommand{\ovi}{O\,{\sc vi}}
\newcommand{\hei}{He\,{\sc i}}  
\newcommand{\heii}{He\,{\sc ii}}

\newcommand{\nev}{Ne\,{\sc v}}  
\newcommand{\hi}{H\,{\sc i}}  
 
\newcommand{\ciii}{C\,{\sc iii}} 
\newcommand{\civ}{C\,{\sc iv}} 
 
\newcommand{\niii}{N\,{\sc iii}} 
\newcommand{\niv}{N\,{\sc iv}} 
\newcommand{\nv}{N\,{\sc v}} 
 
\newcommand{\six}{Si\,{\sc x}} 
 
\newcommand{\mgii}{Mg\,{\sc ii}} 
 
\newcommand{\apj}{ApJ}
\newcommand{\apjs}{ApJS}

\newcommand{\mnras}{MNRAS}

\title{Continuum Shielding and Flow Dynamics in Active Galactic Nuclei}
\author[Doron Chelouche and Hagai Netzer]
       {Doron Chelouche\thanks{email: doron@wise.tau.ac.il; netzer@wise.tau.ac.il} and Hagai
         Netzer\mbox{\raise.9ex\hbox{$\star$}} \\
        School of Physics and Astronomy and the Wise Observatory,
        The Beverly and Raymond Sackler Faculty of Exact Sciences,\\
        Tel Aviv University, Tel Aviv 69978, Israel}

\pubyear{2000}

\begin{document}
\maketitle

\label{firstpage}

\begin{abstract}

We study the ionization, thermal structure, and dynamics of AGN flows which are partially shielded from the central continuum. We utilize detailed non-LTE photoionization and radiative transfer code using exact (non-Sobolev) calculations. We find that shielding has a pronounced effect on the ionization, thermal structure, and the dynamics of such flows. Moderate shielding is especially efficient in accelerating flows to high velocities since it suppresses the ionization level of the gas. The ionization structure of shielded gas tends to be distributed uniformly over a wide range of ionization levels. In such gas, radiation pressure due to trapped line photons can dominate over the thermal gas pressure and have a significant effect on the thermal stability of the flow. Heavily shielded flows are driven mainly by line radiation pressure, and so line locking has a large effect on the flow dynamics. We show that the observed ``$L_\alpha$ ghost'' is a natural outcome in highly ionized flows that are shielded beyond the Lyman limit. We suggest that high velocity AGN flows occupy only a small fraction of the volume and that their density depends only weakly on the velocity field.
\end{abstract}


\section{Introduction}

Gas outflows are common in astrophysical systems in general, and in active galactic nuclei (AGN) in particular. AGN flows span a wide range of ionization levels with spectral signatures extending from the optical to the X-ray band. Direct evidence for outflowing gas in type-I AGN comes from the detection of blueshifted absorption lines, with velocities ranging from a few $\times 100~{\rmn {km~s}^{-1}}$ (so called narrow absorption line, NAL, flows) to more than $10,000~{\rmn {km~s}^{-1}}$ (broad absorption line, BAL, flows). X-ray observations  show that AGN showing  BAL-like features also have relatively weak soft X-ray flux (``soft X-ray weak QSOs'', SXWQ; Brandt, Laor, \& Wills 2000). Among SXWQ, BALQSOs are extremely X-ray weak objects. Recent {\it Chandra} and {\it XMM} observations indicate that soft X-ray weakness results from intrinsic line-of-sight absorption by intervening material (e.g., Gallagher et al. 2002). The properties of the absorbing material  are poorly constrained by current X-ray observations with only a few reliable measurements of the column density ($\gtorder 10^{22}~{\rmn {cm}^{-2}}$; e.g., Green et al. 2001). Similar uncertainties apply also to other flows (e.g., de Kool et al. 2001, Everett, Koenigl, \& Arav 2002; see also Arav et al. 1999). UV NAL outflows and soft X-ray absorption by ionized gas are seen in many low luminosity type-I AGN (George et al. 1998, Crenshaw et al. 1999) and a possible physical connection between the two has been suggested by Mathur et al. (1994). 

Like stellar winds, AGN outflows are thought to be driven by radiation pressure force (e.g., Arav, Li, \& Begelman 1994, Koenigl \& Kartje 1994, Murray, Grossman, Chiang, \& Voit 1995, Proga, Stone, \& Kallman 2001, and Chelouche \& Netzer 2001; hereafter CN01, but see Begelman, de Kool, \& Sikora 1991 for a different model). However, the differences in the geometry, the distance scales, and the underlying continuum between the groups are very large. A long standing issue concerns the fact that BAL flows have a relatively uniform ionization structure over a wide range of velocities. This can be explained theoretically if the flow is kept from becoming too dilute by means of a confining medium (e.g., Arav, Li, \& Begelman 1994; but see Krolik 1979 and Mathews \& Blumenthal 1977) or by means of continuum shielding (Murray et al. 1995, Proga et al. 2000). The latter is the focus of this paper.

This work is part of a series of papers investigating the ionization and dynamics of radiation pressure driven flows in AGN.  In the first paper (Chelouche \& Netzer 2003; hereafter CN03) we have developed a general non-Sobolev scheme for calculating the ionization and dynamics of AGN flows. Here we focus on the physics of shielded flows, i.e., those flows which are illuminated by a modified (absorbed)  type-I AGN continuum.  Specifically, we attempt to answer the following key questions: 1) What are the dominant ionization levels in shielded flows? 2) What governs the dynamics of shielded flows? 3) What is the preferred geometry of high velocity flows? 4) Can self-consistent photoionization and dynamical calculations account for the observed properties of BAL flows?

This paper is organized as follows: In section 2 we present the general method and the basic assumptions. Results pertaining to the ionization level, thermal structure, and radiation pressure  are given in section 3. The radiation pressure force is discussed in section 4. In section 5 we calculate the structure and dynamics of shielded flows, focusing on their velocity profiles and spectral features. The conclusions and summary are presented in section 6.

\section{Method}

\begin{figure}
\includegraphics[width=84mm]{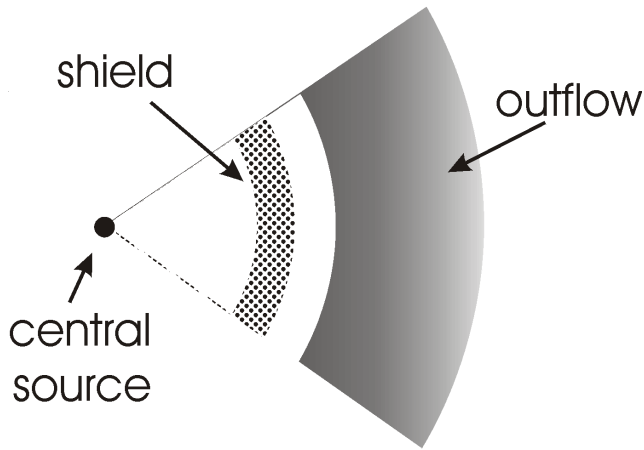} 
\caption{The flow--shield geometry. The shield is situated between the continuum source and the flow and may or may not be coupled, mechanically, to the flow. The ionizing flux impinged upon the flow is due to the transmitted continuum. It depends on $U_x^{\rmn {shield}}$ and $N_H^{\rmn {shield}}$ (see text). Spherical geometry is assumed.}
\label{geometry}
\end{figure}

Our aim is to study the effect of shielding on the dynamics, ionization, and thermal structure of radiation pressure driven photoionized flows in AGN. The motivation stems from observations as well as from previous theoretical studies which suggest the possible effect of shielding on the gas dynamics. Throughout this work we assume flows that are exposed to a modified type-I AGN ionizing continuum whose properties depend on shielding material (hereafter ``the shield'') which lies between the ionizing source and the base of the flow. The shield--flow geometry is depicted in figure \ref{geometry}. 

\subsection{Self-consistent calculation of the flow structure and dynamics}

This work focuses on the properties of radiation pressure driven flows as described in CN03. Below is a brief outline of the formalism and the reader is referred to CN03 for a more detailed description.

The equation of motion is
\begin{equation}
v\frac{dv}{dr}= \frac{1}{\rho} \left ( F_{\rmn {rad}} -\frac{dP_{\rmn {tot}}}{dr} \right ) - g(r),
\label{eqnmot}
\end{equation}
where $v$ is the velocity, $r$ the radial coordinate, $\rho$ the gas density, $F_{\rmn {rad}}$ the radiation pressure force, $P_{\rmn {tot}}$  the total pressure (the sum of gas and internal radiation pressure), and $g$ the gravity. Both $F_{\rmn {rad}}$ and $P_{\rmn {tot}}$ depend on the ionization and thermal structure of the flow, which in turn depends on the ionization parameter, $U_x$, (the ratio of the photon density in the range 0.1-10\,keV and the hydrogen number density), and the  spectral energy distribution (SED) of the ionizing flux. In all calculations we assume low density flows ($<10^{12}~{\rmn {cm}^{-3}}$). Thus, collisional processes are negligible (CN01). Unless otherwise stated, we assume the following continuity condition,
\begin{equation}
\rho \propto r^{-2} v^{-1}.
\label{sph_cont}
\end{equation}
The local radiation pressure force is calculated  using the force multiplier formalism,
\begin{equation}
F_{\rmn {rad}}(r)=H_cM(r),
\label{frad}
\end{equation}
where
\begin{equation}
H_c={\rmn {max}} \left ( n_H,n_e \right ) \frac{\sigma_T L_{\rmn {tot}}}{ 4\pi r^2 c},
\label{hc}
\end{equation}
$\sigma_T$ is the Thomson cross-section, $n_H$ the hydrogen number
density, $n_e$ the free electron number density, and $L_{\rmn {tot}}$ the
bolometric luminosity of the source.  Note that by choosing ${\rmn {max}}
\left ( n_H,n_e \right )$, instead of $n_e$, we allow for the Compton radiation pressure force in both ionized and neutral gas.

The force multiplier, $M(r)$,  includes the contribution of  all scattering and absorption processes: bound-bound - $M_{\rmn {bb}}$, bound-free - $M_{\rmn {bf}}$, and Compton scattering. Another useful parameter that has been used extensively in older, Sobolev-type calculations is the local electron scattering optical depth, 
\begin{equation}
t={\rmn {max}} \left ( n_H,n_e \right ) \sigma_T v_s \left ( \frac{dv}{dr} \right )^{-1},
\label{t}
\end{equation}
where $v_s$ is the sound speed. This parameter converges to the usual definition of $t$ that includes $n_e$ instead of $n_H$ for ionized gas and is  suitable for both highly ionized and neutral flows. 

The total pressure is the sum of gas pressure, $P_{\rmn {gas}}$ and internal radiation pressure, $P_{\rmn {rad}}$, due to scattering of resonance line photons (Elitzur \& Ferland 1986).  We calculate $P_{\rmn {rad}}$ according to the scheme outlined in CN03 which is suitable for differentially expanding flows with a finite optical depth. 

The above  scheme is used to obtain a self-consistent solution for the flow structure and dynamics given a set of initial conditions, $r_0,~U_x(r_0)$, $L_{\rmn {tot}}$,  the initial velocity, $v_0$, and the mass of the central object. We iterate between the photoionization and thermal equations, the radiation pressure force calculations, the equation of motion, and the continuity condition, until the density profile converges to less than $1\%$. Using this method, the mass flow rate is determined by the initial conditions and not from critical point analysis (see CN03).

\subsection{The modified ionizing continuum}

An important input for the calculations is the assumed SED. X-ray observations of SXWQ cannot constrain the ionization level of the intervening absorber. Signatures of highly ionized absorbing gas are common in type-I objects and recent works (e.g., Proga, Stone, \& Kallman 2001) show that this may also be the case in SXWQ. In this work we  model the shield as a uniform slab whose properties are defined by its ionization parameter, $U_x^{\rmn {shield}}$ and column density, $N_H^{\rmn {shield}}$.  This allows us to study a wide range of shield properties. We assume that the shield completely covers the ionizing source and is static (i.e., absorption lines with thermal widths).

The shield is exposed to a typical type-I AGN continuum with a UV
bump, $\alpha_x=0.9$ and $\alpha_{ox}=1.4$ ($L_E\propto E^{-\alpha}$,
see CN01 and references therein). We calculate the self-consistent
ionization and thermal structure of the shield using {\sc ion\,2002},
the 2002 version of the  photoionization code {\sc ion} (Netzer 1996), and obtain the transmitted spectrum through the shield, ignoring emission features.

Our thermal calculations include the effect of adiabatic cooling which depends on $dv/dr$ and $v/r$. We find that, in general, adiabatic cooling can be neglected for sub-relativistic flows close to the AGN ($r <100\,$pc). Flows which lie at larger distances are probably not in photoionization equilibrium and shall not be treated in this work.

\begin{figure}
\includegraphics[width=84mm]{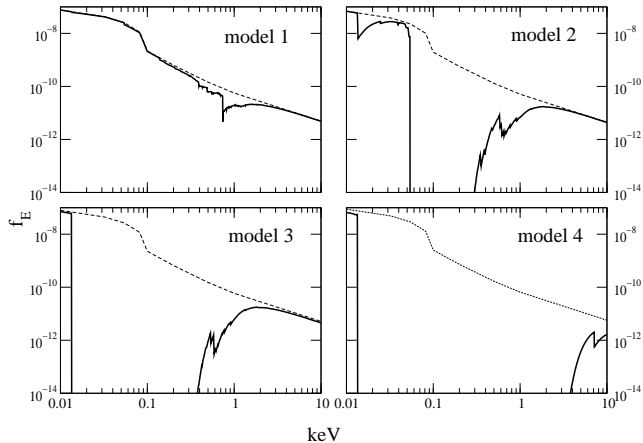} 
\caption{The four shielded continua used in this work.  Note the heavy X-ray absorption of model 4 where the continuum recovers only beyond $\sim 10$\,keV. Model 1 has typical warm absorption features (\ovii~ and \oviii~ edges). The dashed line marks the intrinsic type-I SED used in this work.}
\label{sdef}
\end{figure}

\begin{table} 
\centering 
\begin{tabular}{cccl} 
\hline 
No.  & $U_x^{\rmn {shield}}$ &  $N_H^{\rmn {shield}}$ & comments \\ 
&  &  $(\rmn {{cm}}^{-2})$ & \\ 
\hline
1 &  $10^{-1}$ & $10^{22}$ & warm absorption \\
2 &  $10^{-2.5}$  & $10^{22}$ & optically thick \heii~ edge \\
3 &  $10^{-3}$ & $10^{22}$ & optically thick \hi~ \& \heii~ edges \\
4 &  $10^{-3}$ & $10^{24}$ & extreme neutral shielding \\
\end{tabular}
\caption{Definition of the shielding models used in this work.} 
\label{tdef}
\end{table}

We define four types of modified continua that are shown in figure \ref{sdef} and are defined in table \ref{tdef}.  Model 1 has prominent ``warm absorption'' features due to \ovii~ and \oviii~ edges. Model 2 is optically thick beyond the \heii~ edge, and recovers above $\sim$1\,keV. Models 3 and 4 are optically thick above the \hi~ edge and recover at $\sim$1\,keV and $\sim$10\,keV, respectively.  We have also explored a set of neutral shields that are commonly used in X-ray spectral modelling (e.g., the {\sc wabs} model in {\sc xspec}; e.g., George et al. 1998). 

In our shielded flow models, $P_{\rmn {tot}}$ and $F_{\rmn {rad}}$ depend on $U_x$ and the SED. In order to allow a meaningful comparison between flows that are exposed to different continua, we have changed the definition of $U_x$ to denote  {\it the value of the ionization parameter which would be defined  for  unshielded gas at the same location and with the same density}.

\section{Ionization, temperature and pressure of shielded outflowing gas}

\subsection{General considerations for neutral shields}

\begin{figure}
\includegraphics[width=84mm]{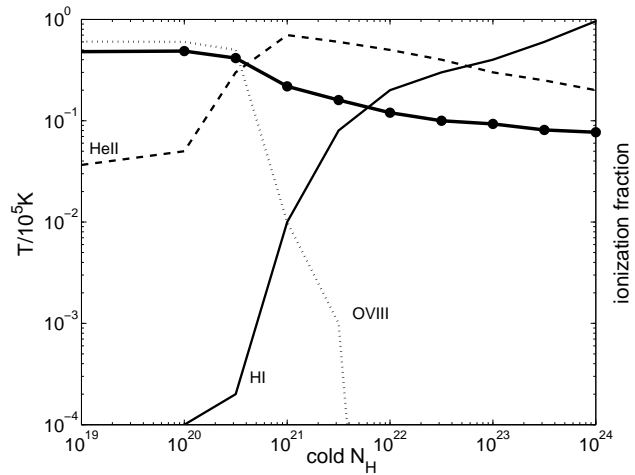}
\caption{The temperature (in units of $10^5$K) and ionization structure of \hi, \heii, and \oviii~ in shielded gas with $U_x=0.1$ as a function of $N_H^{\rmn {shield}}$ for shielding by neutral material. Note the sharp drop in temperature and in \oviii~ fraction for $N_H^{\rmn {shield}} \gtorder 10^{21}~{\rmn {cm}^{-2}}$. For $N_H^{\rmn {shield}} \gtorder 10^{22}~{\rmn {cm}^{-2}}$, the gas is almost completely neutral.}
\label{colds}
\end{figure}

We first study the effect of neutral shielding assumed to be most relevant to SXWQ.  Figure \ref{colds} shows the dependence of the ionization structure of \hi, \heii~ and \oviii~ for shielded gas with $U_x=0.1$ on the properties of the  neutral shield. As shown, the level of ionization of such gas drops with increasing $N_H^{\rmn {shield}}$. For $N_H^{\rmn {shield}} \sim 10^{22}~{\rmn {cm}^{-2}}$, the \oviii~ abundance drops by almost two orders of magnitude and the ionization level resembles that of the broad line region (BLR). The temperature traces the ionization level of the flow and decreases by a factor $\sim 4$ from small column density shields to the $N_H^{\rmn {shield}}=10^{22}~{\rmn {cm}^{-2}}$ shield. 

\subsection{Thermal structure and stability}

\begin{figure}
\includegraphics[width=84mm]{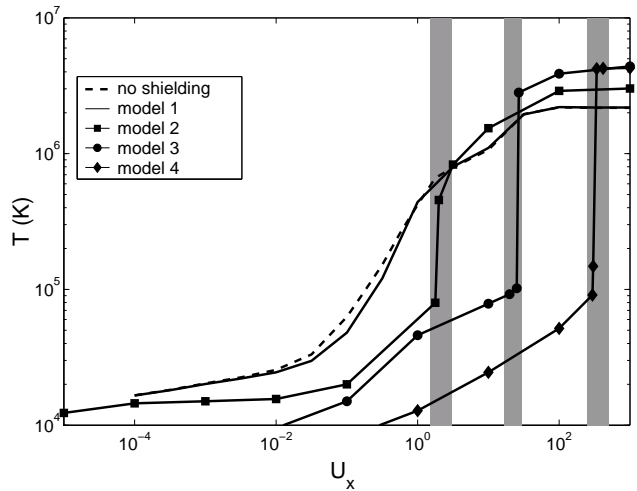} 
\caption{The temperature of photoionized gas as a function of $U_x$ for various shielding models. As shielding increases, the temperature is reduced and higher temperatures require larger $U_x$. Note the sharp transition to high temperatures designated by the shaded areas. The dashed curve corresponds to unshielded gas.}
\label{tshield}
\end{figure}

The temperature of the shielded gas as a function of $U_x$ is shown in figure \ref{tshield} for all four models.
Warm absorption-type shield (model 1) has only a small and localized effect on the ionization and the temperature of the shielded gas. When shielding increases (model 2), the reduction in electron temperature is more pronounced due to the large deficit in ionizing photons. However, above some critical $U_x$ ($U_x>1$ for model 2), the temperature climbs quickly to the Compton temperature which is slightly above the temperature of the unshielded gas due to the different mean photon energy. Shielding above the \hi~ edge (model 3) suppresses the temperature up to $U_x \simeq 10$. In this case, the shielded gas is neutral up to $U_x=10^{-2}$.  For the largest shielding (model 4), the gas is almost neutral up to $U_x\sim 10^2$. Thus, heavily shielded gas that shows absorption features due to \ovii~ and \oviii~ edges is $\sim 10^3$(!) times more dilute than non-shielded gas at the same location showing the same features. 

Figure \ref{tshield} shows also the presence of a critical ionization parameter, $U_x^{\rmn {crit}.}$ (marked by shaded areas) for the shielded gas, beyond which the gas temperature climbs to the Compton temperature. Heavier shielding results in large $U_x^{\rmn {crit}.}$. The reason for the sharp rise in temperature over a narrow $U_x$ range ($\Delta U_x/U_x^{\rmn {crit}.} \ll 1$) is the peculiar ionization structure of \heii. When \heii~ abundance is large, the temperature is low so recombination cooling is effective and the gas remains at low temperatures (see also Stevens 1991). Once $U_x^{\rmn {crit}.}$ is reached, \heii~ ionizes, the temperature rises, recombination becomes less effective, most metals ionize and the cooling rate decreases sharply. Under these conditions, the gas reaches its Compton temperature. 

While photoelectric absorption from excited levels is negligible for unshielded non-LTE gas, this is not the case for heavily shielded, partly ionized gas (e.g., models 3 and 4) with large $U_x$. Under these conditions, continuum pumping can populate the excited levels in regions of small optical depth.

\begin{figure}
\includegraphics[width=84mm]{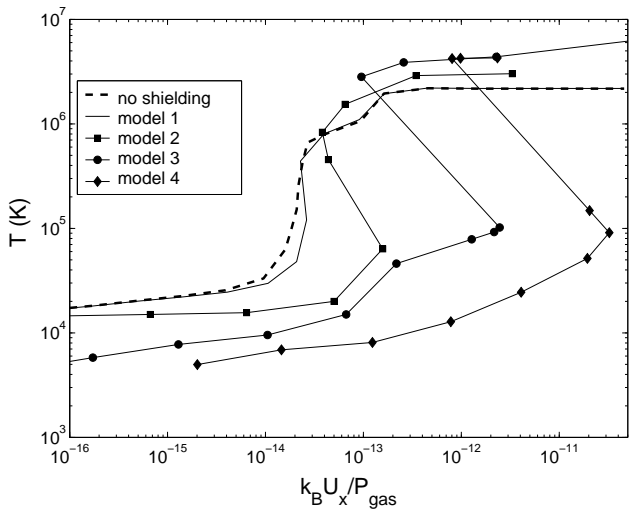}
\caption{ Thermal stability curves for  various shielded flows. The instability regions (the sections with negative derivatives) occupy only a small range in $U_x$ (the shaded areas in figure \ref{tshield}). The dashed curve corresponds to an unshielded case. ($k_B$ is the Boltzmann constant).}
\label{tstab}
\end{figure}

Shielding and internal radiation pressure can have a profound effect on the thermal stability curve of photoionized, shielded gas. We first show the effect of shielding alone and then incorporate the effects of internal radiation pressure. 

Shielding of photoionized gas enhances and extends its thermal instability (the part of the curve with $dT/d(U_x/P_{\rmn {gas}}) <0$) due to the suppression of the UV flux. This is illustrated in figure \ref{tstab}. Given our intrinsic SED, non-shielded gas is thermally stable over the entire range of $U_x$. Model 1 shielding results in a small region of marginal stability (see also Marshall et al. 1993, Krolik \& Kriss 2002). Increased shielding results in appearance of larger instability regions. These regions correspond to a narrow range in $U_x$ ($\Delta U_x/U_x^{\rmn {crit}.} \ll 1$; the shaded areas in figure \ref{tshield}). The shape of the stability curves for heavily shielded flows is relatively independent on the shape of the intrinsic SED. 

\begin{figure}
\includegraphics[width=84mm]{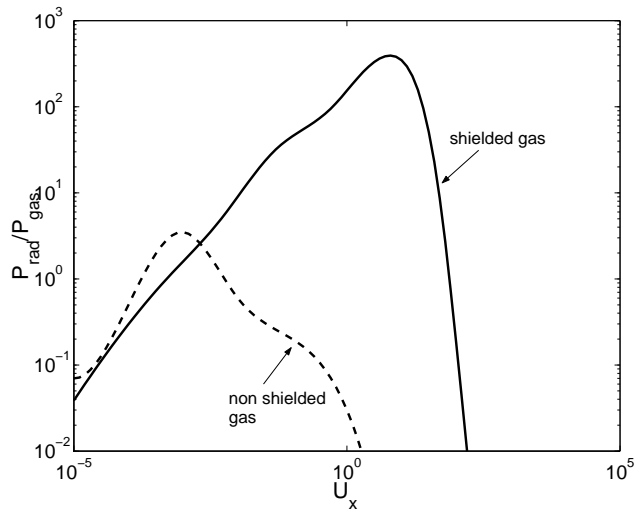} 
\caption{The ratio of radiation to gas pressure as a function of $U_x$ for model 3 shielding and shielded gas which is optically thick to resonance lines (see text). For highly ionized shielded gas ($U_x\simeq 1$) radiation pressure dominates the total pressure. For low $U_x$, $P_{\rmn {gas}} \gg P_{\rmn {rad}}$ due to the low ratio of photon to gas density. When $U_x$ is large, the gas is fully ionized and scattering of line photons is negligible. Note the large difference in the contribution of $P_{\rmn {rad}}$ to the total pressure between shielded and non-shielded gas.}
\label{ptot}
\end{figure}

We considered in detail the effect of $P_{\rmn {rad}}$ on the thermal stability of the gas. The exact value of $P_{\rmn {rad}}$ depends on the global properties of the flow (temperature, ionization structure and dynamics), and  on the shield--flow configuration. To simplify the problem and to obtain an upper limit on this effect, we focus on the extreme case where the base of the flow and the shield are comoving with no differential motion. We assume that the flow has a uniform ionization structure and is optically thick to resonance line absorption (in particular $L_\alpha$, \civ~$\lambda 1549$, \heii~$\lambda 304$). This results in the largest possible $P_{\rmn {rad}}$ (see figures 7,\,8 in CN03). In figure \ref{ptot} we show the contribution of $P_{\rmn {rad}}$ to $P_{\rmn {tot}}$. $P_{\rmn {rad}}$ is  larger than $P_{\rmn {gas}}$ by more than two orders of magnitude for highly ionized shielded gas with substantial line opacity. This result is very different from the typical values obtained for non-shielded gas (CN03) for the following reasons: For a given $U_x$, the level of ionization of a shielded gas is always lower than the ionization level of a non-shielded gas. This means more optically thick lines and , hence, larger $P_{\rmn {rad}}$. For very large $U_x$, even the shielded gas is ionized and $P_{\rmn {rad}}$ drops accordingly.

\begin{figure}
\includegraphics[width=84mm]{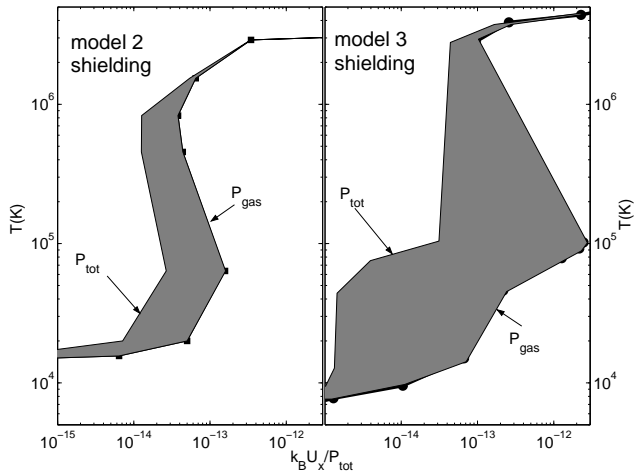} 
\caption{ The effect of line pressure on the stability curve for model 2 (left panel) and  model 3 (right panel) shielding. Both diagrams refer to a static configuration (see text). The stability curves of all flows whose opacity is small compared to the shield opacity lie within the shaded area. Some of those curves can be thermally stable for the entire range of $U_x$.} 
\label{stab}
\end{figure}

Isobaric perturbations lie, by definition, along a $P_{\rmn
  {tot}}=$\,const. curve (see Bottorff et al. 2001 for  instructive
discussion on stability curves). We therefore  carried the same
stability analysis substituting $P_{\rmn {tot}}$ for $P_{\rmn {gas}}$ in the
abscissa of figure \ref{tstab}. The new stability curves for model 2
and 3 shielding are shown in figure \ref{stab}. The inclusion of
$P_{\rmn {rad}}$ significantly alters the shape of the stability curve
and narrows the range of ionization parameter where thermal
instabilities occur. These newly calculated stability
curves illustrate the maximum deviation from a $P_{\rmn {tot}}=P_{\rmn
  {gas}}$ stability curve since dynamical effects within the flow and
between the flow and the shield have not been included. Specific flow
models would therefore lie within the shaded 
area bounded by the two extreme stability curves (including and
excluding the maximum $P_{\rmn {rad}}$). Figure \ref{stab} shows
that it is possible, in principal, to have a flow configuration which
is thermally stable for all values of $U_x$. This stabilizing effect
is especially important for low velocity, optically thick flows. We
note that $P_{\rmn {rad}}$ cannot suppress isobaric perturbation with
wavelength $\lambda_p$ ($\delta T \propto e^{i2\pi r/\lambda_p}$)
shorter than the mean free path of the photon in the gas. This is the
physical length corresponding to a line optical depth of $\sim 1$.

\subsection{The ionization structure}

\begin{figure*}
\includegraphics[width=178mm]{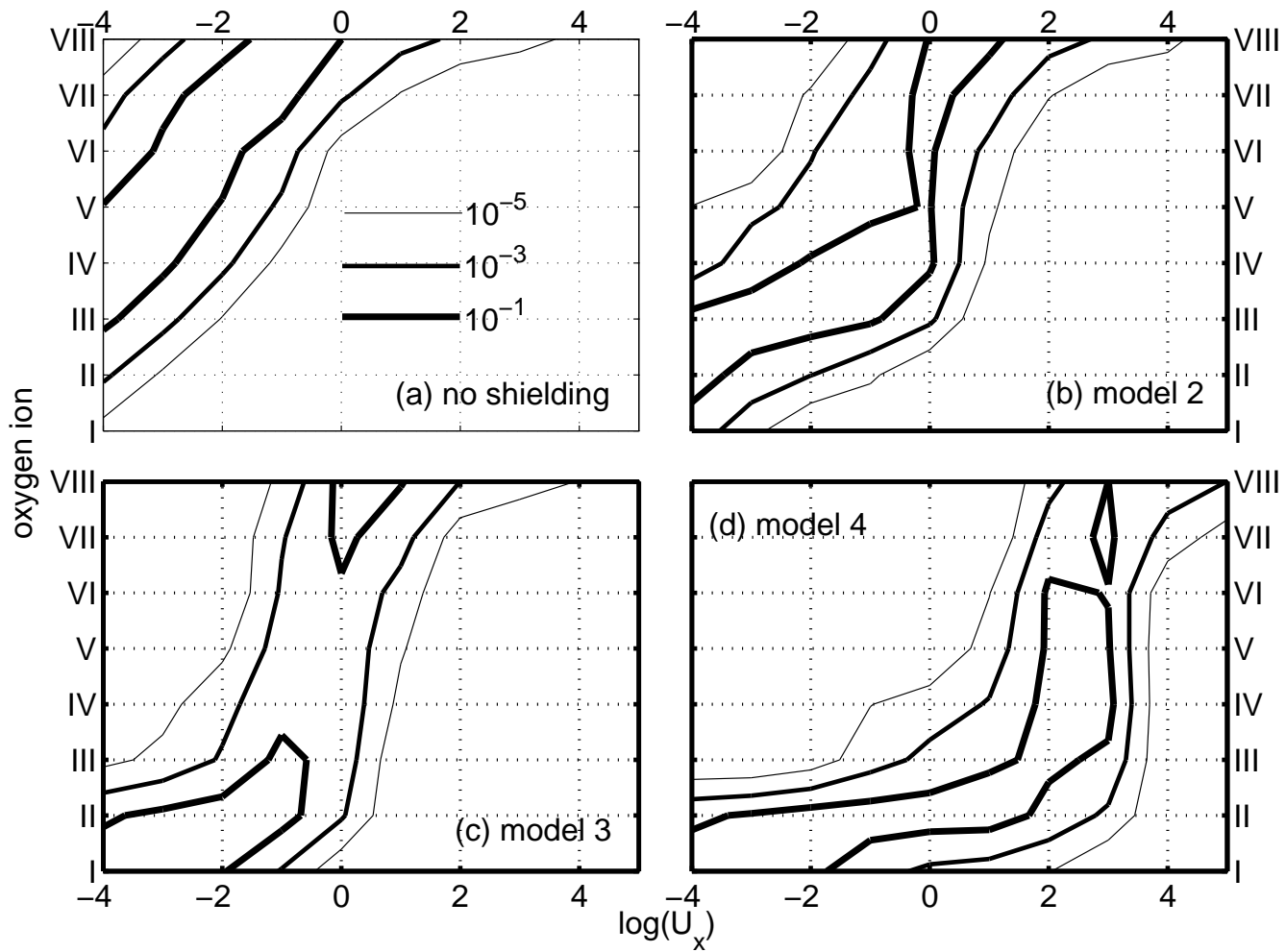}
\caption{The fractional ionization of \oi--\oviii~ as a function of $U_x$ for gas which is exposed to (a) a non-shielded ionizing continuum, (b) model 2, (c) model 3, and (d) model 4 shielding. Each contour level (marked by a different line) represents two orders of magnitude in abundance ($10^{-5},~10^{-3},~10^{-1}$). As shielding increases, the neutral ionization structure extends to larger $U_x$ and more oxygen ions have similar abundances at high ionization parameters (e.g., for model 4 shielding, all ions from \oiii~ to \oviii~ show the same abundances at $U_x\sim100$).}
\label{ofrac}
\end{figure*}

Next we study the effect of shielding on the ionization structure of the gas. We demonstrate this by showing the oxygen ionization structure. This is shown in figure \ref{ofrac}. As expected, the deviations of the ionization structure from the unshielded case are more pronounced for heavier shielding. For example, \oiii~ is very abundant up to $U_x < 10^{-3}$ for unshielded gas, but remains abundant up to $U_x=100$ for model 4 shielding.  

Shielding has another important effect on the ionization structure of the gas. For non shielded gas, the ionization structure of  oxygen is dominated by the three most abundant ions, each with  an ionization fraction of $\sim 0.3$. This is shown, for example, in figure \ref{ofrac}a for $U_x = 0.01$ where most oxygen is \ov, \ovi~, and \ovii. As shielding increases, the distribution of ionization levels is wider (see the knee which develops in figures \ref{ofrac}b-d as the contours become vertical at high $U_x$). A most extreme example is that of figure\, \ref{ofrac}d where for $U_x\simeq 100$, the oxygen ionization structure is very uniform with \oiii--\oviii~ all  having similar fractional ionizations of $\sim 0.1$. The reason for this peculiar ionization structure is the shape of the modified SED which shields \oii~ and higher ionization levels from the ionizing UV photons, but allows X-ray photons to penetrate the gas and cause inner shell ionization. The density of the X-ray photons above the K ionization edge is similar for all oxygen ions and the distribution among the various ions is more uniform. Thus, heavily shielded gas may show a myriad of ionization levels all coming from the same location. This  may explain the wide range of ionization levels observed in BAL flows. 

\section{The radiation pressure force}

Here we study the effect of shielding on the radiation pressure force. We focus on model\,2--4 shielding since model\,1 shielding has only a small effect.  In the following diagrams the value of $M$ is normalized to the Compton radiation pressure due to the non-shielded continuum in order to allow comparison between different shields.

\begin{figure}
\includegraphics[width=84mm]{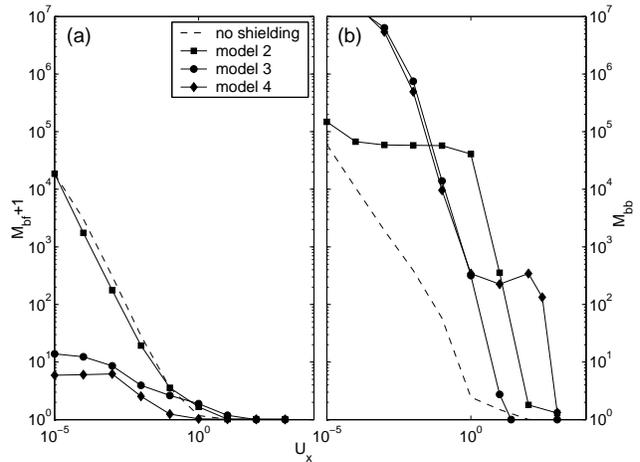} 
\caption{The effect of shielding on the radiation pressure force (a)
  due to continuum processes ($M_{\rmn {bf}}$) and (b) due to lines ($M_{\rmn {bb}}$). Note the decline of the radiation pressure force with $U_x$ for all models. Note also the increase of $M_{\rmn {bb}}$ and the decrease of $M_{\rmn {bf}}$ with shielding (see text). The results for non-shielded cases are shown for comparison (dashed line).}
\label{mthick}
\end{figure}

Results for the bound-free ($M_{\rmn {bf}}$) and bound-bound ($M_{\rmn {bb}}$) force multiplier are shown in figure \ref{mthick}a and \ref{mthick}b.  As shielding increases, $M_{\rmn {bf}}$ decreases due to the growing deficit in UV photons that carry most of the momentum. This is more pronounced for low $U_x$ and heavy shielding (model\,3 and 4) where the largest contributions to $M_{\rmn {bf}}$ are due to absorption by \hi~ and \heii. Contrary to $M_{\rmn {bf}}$, $M_{\rmn {bb}}$ increases with shielding, especially for low $U_x$, since the main driving lines of low ionization species lie below the \heii~ and \hi~ edges, and are thus exposed to the unshielded continuum.

\begin{figure}
\includegraphics[width=84mm]{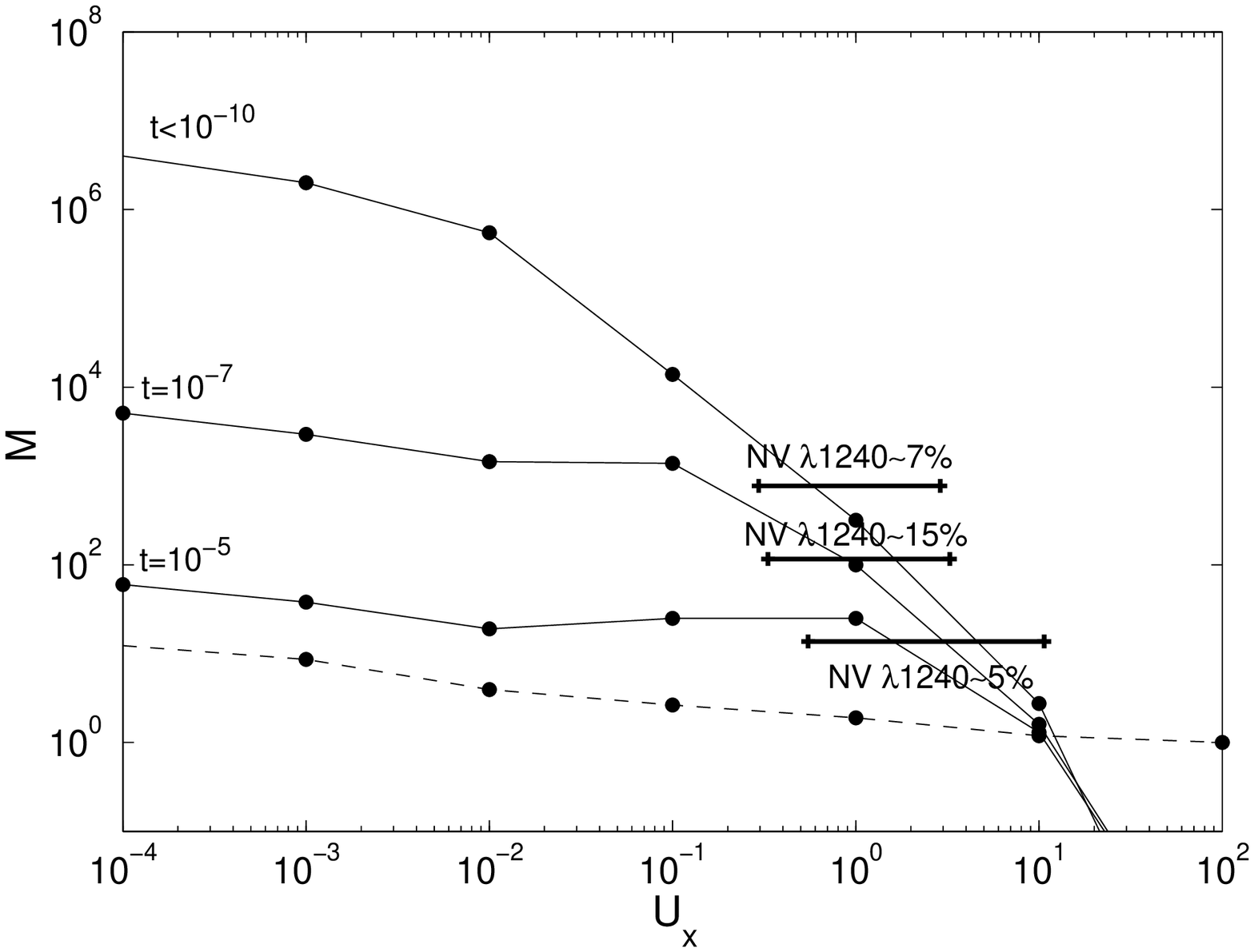}
\caption{ $M_{\rmn {bb}}$ (solid line) and $M_{\rmn {bf}}$ (dashed line) for various values of $t$ for model 3 shielding. Note the rapid decline of $M_{\rmn {bb}}$ with $t$.  The range of $U_x$ where \nv~$\lambda 1240$ contributes significantly to $M$ is marked next to each curve. As shown, \nv~$\lambda 1240$ contribution is significant for ionized shielded flows.}
\label{t_thick}
\end{figure}

Next we study the effect of dynamics on the radiation pressure force. We do so by using the Sobolev $t$-parameterization (equation \ref{t}). This implicitly assumes the Sobolev approximation, i.e., ignoring line locking and line blanketing and discontinuities in the flow. As shown in figure \ref{t_thick}, dynamics has a substantial effect on $M_{\rmn {bb}}$ but a negligible effect on $M_{\rmn {bf}}$. $M_{\rmn {bb}}$ declines with increasing $t$ since many lines become optically thick and their contribution to $M$ declines. This is most pronounced for models\,3 and 4 and for low $U_x$ where the largest contribution to $M_{\rmn {bb}}$ are due to \hi~ and \hei~ absorption lines.  Our calculations show that moderate shielding results in more efficient acceleration for large $U_x$ and $t$ due to the larger contribution of metal lines which become optically thick at a significantly larger $t$.

\begin{table}
\centering 
\begin{tabular}{lcc} 
\hline 
Line      &    $\tau_l$ &  $M_l$ \\ 
\hline 
& model 2 & \\
& $U_x=1$ & ($M_{\rmn {bb}}=5$) \\
\hline
\nev~ $\lambda 359$ & 2.2 & 0.15 \\
\civ~ $\lambda 245$ & 1.5 & 0.14 \\
\niv~ $\lambda 247$ & 2.3 & 0.13 \\
\hi~ $\lambda 1217$ & 0.18 & 0.11 \\
\six~ $\lambda 276$ & 1.1 & 0.11 \\
\civ~ $\lambda 1549$ & 44 & 0.04 \\
\hline
& model 3 & \\
& $U_x=8$ & ($M_{\rmn {bb}}=3.3$)\\
\hline
\hi~ $\lambda 1217$ & 1.3 & 0.6 \\
\hi~ $\lambda 1027$ & 0.2 & 0.2 \\
\niii~ $\lambda 991$ & 0.5 & 0.16 \\
\nv~ $\lambda 1238$ & 5 & 0.13 \\
\ciii~ $\lambda 978$ &  20 & 0.1 \\
\civ~ $\lambda 1549$ & 16 & 0.07 \\
\hline
\end{tabular}
\caption{Contribution of the most important lines to the force 
  multiplier assuming $t=10^{-4}$. Note that for model 3 shielding,
  only near UV and optical lines participate in driving the gas since the continuum is devoid of far UV photons. The gas temperature in both cases is $4\times10^4$K.}
\label{t:lines_thick}
\end{table}

Table \ref{t:lines_thick} lists the important lines that drive the flow for  model 2 and 3 shielding and the case of $t=10^{-4}$. We chose the value of $U_x$ such that the temperature is similar in the two cases. This means an order of magnitude difference in $U_x$ because of the very different shielding.  For moderate shielding, the contribution of lines near the UV bump (around $250{\mbox{\AA}}$) is the largest. For heavy shielding, near UV and optical lines are the largest contributors to $M_{\rmn {bb}}$ since the flux above the \hi~ edge is negligible. X-ray lines also contribute to $M_{\rmn {bb}}$ since the continuum recovers at around 2\,keV. In all shielded flows, the main driving lines are  very different from the driving lines in non-shielded cases and fewer lines contribute significantly to $M_{\rmn {bb}}$ (see table 1 in CN03). The contribution of absorption from excited levels is negligible for such optically thick gas.

CN03 argued that the contribution of line locking to the total radiative acceleration is small in non-shielded AGN gas. Nevertheless, observations of \civ~$\lambda 1549$ show that line locking in general, and that of $L_\alpha$ on \nv~$\lambda 1240$ in particular, can have a significant and observable effect on the dynamics of BAL flows (the so called ``$L_\alpha$ ghost phenomenon''; e.g., Korista et al. 1996, Arav et al. 1996). Arav (1996) argued that the effect of line locking can be important if optical and near UV lines are solely responsible for accelerating the gas. Below we discuss two issues related to line-locking: self-line locking where absorption lines are exposed to the peak of their corresponding emission lines, and $L_\alpha$-\nv~$\lambda 1240$ line locking.

\begin{figure}
\includegraphics[width=84mm]{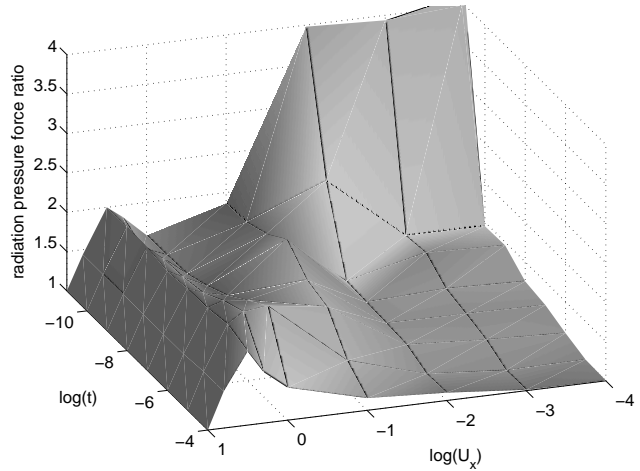} 
\caption{The effect of line locking on the radiation pressure force for model 3 shielded gas. The ratio of radiation pressure force when line locking is accounted for to the case where it is neglected is shown as a function of $U_x$ and $t$. Emission line properties were taken from Tefler et al. (2002). The effect of $L_\alpha$ locking becomes important for optically thin, low ionization flows and diminishes rapidly with $t$ as $L_\alpha$ becomes optically thick and with $U_x$ due to decreased abundance of \hi.  The contribution of \civ~$\lambda 1549$, \nv~$\lambda 1240$, \ovi~$\lambda 1035$ is significant along the ridge at $U_x\gtorder 1$ (see text).}
\label{mlock}
\end{figure}

The effect of self locking for model 3 shielded flow is shown in figure \ref{mlock}. The diagram shows the ratio of $M$ calculated with and without self locking as a function of $U_x$ and $t$. The lines that are included are  \mgii~$\lambda 2799$, $L_\alpha$, \nv~$\lambda1240$, and \ovi~$\lambda 1035$ with properties as given in Telfer et al. (2002). We assume that emission and absorption line profiles have no relative shifts, i.e., maximum line locking. Comparing to the non-shielded case (figure 6 in CN03), we note that shielding increases the effect of line locking since, in this case, the number of lines that contribute significantly to $M_{\rmn {bb}}$  is smaller. The contribution of $L_\alpha-L_\alpha$ locking is most pronounced for low ionization, optically thin gas, and can increase $M$ by a large factor. This contribution is significantly smaller for larger $U_x$ because of the drop in the \hi~ abundance, and for large $t$ ($t>10^{-10}$) when $L_\alpha$ becomes optically thick. A most peculiar feature is a ridge with an almost constant ratio of $\sim 2$, extending over the entire range of $t$ for $U_x\simeq 1$. This is naturally explained by figure \ref{ofrac} where, for a narrow range of $U_x$, many ionization levels have similar abundances. Line locking is, in general, less important for less shielding where the UV flux is unattenuated. 

The  contribution of \nv~$\lambda 1240$ locking to the total radiation pressure force is of great importance in the study of BAL flows.  CN03 have shown that the contribution of \nv~$\lambda 1240$ is $\sim 1\%$ for non-shielded, solar metallicity gas. We have recalculated this contribution for shielded flows and show the results in  figure \ref{t_thick}. The contribution of \nv~$\lambda 1240$ to $M$ remains small as long as the UV bump contributes considerably to the total flux. Larger shielding extinguishes the UV flux above the \hi~ edge and increases the relative contribution of this line. The total contribution to $M$, for model 3 shielding, is $\sim 10\%$ for a wide range of $t$ and $U_x$. Thus, for model 3 shielding and typical $L_\alpha$ emission line, the radiation pressure force due to \nv~$\lambda 1240$ is larger than $50\%$ of the total force (note that in figure \ref{t_thick}, the contribution of \nv~$\lambda1240$ was calculated relative to the continuum level). For model 4 shielding, the relative contribution of \nv~$\lambda 1240$ declines due to the lower fractional abundance of \nv~ as a result of the more uniformly distributed ionization structure. 

\section{The dynamics of shielded flows}

We calculated several self-consistent dynamical flow models in order to answer some of the key questions raised in section 1. These include the effect of shielding on the flow dynamics and a qualitative comparison of our model predictions to UV and X-ray observations of BALQSOs. The calculations make use of the non-Sobolev formalism described in CN03. In all subsequent calculations we assume $L_{\rmn {tot}}=10^{45}~{\rmn {erg~s}^{-1}}$ intrinsic type-I AGN continuum, and neglect gravity relative to $F_{\rmn {rad}}$.

\subsection{The effect of shielding on the flow velocity}

\begin{figure}
\includegraphics[width=84mm]{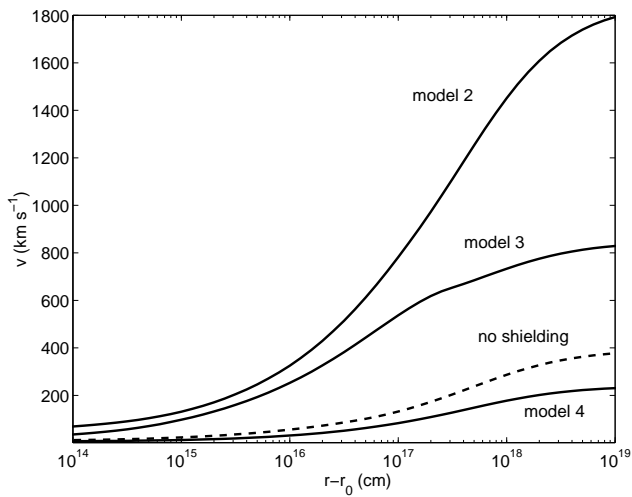}
\caption{The effect of different types of shield on the dynamics of a flow with given initial conditions (see text). Moderate shielding tends to increase the terminal velocity of the flow. Heavy  shielding lowers the terminal  velocity.} 
\label{vshield}
\end{figure}

We first study the effect of shielding on the dynamics of a flow launched from $r_0=10^{18}~{\rmn {cm}}$ and characterized by $U_x(r_0)=10^{-2}$. The velocity profile of shielded and unshielded flows with such properties are shown in in figure \ref{vshield}. The most important conclusion  is that shielded flows can be accelerated to much larger velocities. Moderate shielding (model 2) suppresses the ionization level and maintains a large radiation pressure force due to the intense continuum below the \heii~ edge. This drives the flow to higher velocities compared to more ionized unshielded flows (factor 4 difference in the terminal velocity).  Model 3 shielding results in a lower ionization flow which is driven mainly by \hi~ and \heii~ lines. Such lines become optically thick already at small $t$. This and the large deficit in UV driving photons result in a lower velocity flow (factor $\sim$2 lower than model 2 shielding). Even heavier shielding results in an almost neutral flow whose lines are optically thick and cannot effectively drive the flow to high velocities.  

Next we check whether our dynamical model can explain the properties of BALQSOs. Such flows are characterized by  UV and X-ray column densities of $\sim 10^{22}~{\rmn {cm}^{-2}}$ (Arav, Korista, \& de Kool 2002, Gallagher et al. 1999, 2001, 2002) and  typical BAL velocities of order $10,000~{\rmn {km~s}^{-1}}$. This is done in the context of spherically expanding continuous flows (equation \ref{sph_cont}) where $U_x \propto v$.  Consider model 2 shielding and note that in order for the flow to show UV absorption features, its ionization parameter must be smaller than $10$ (otherwise it would be fully ionized, see figure \ref{tshield}). Thus, if the BAL trough extends to $\sim 10^4~{\rmn {km~s}^{-1}}$ then the entire observed part of the flow must have $U_x<10$. Given the $U_x \propto v$ dependence, the part of the flow that is traveling with $\sim 1000~{\rmn {km~s}^{-1}}$ would have $U_x<1$. 

\begin{figure}
\includegraphics[width=84mm]{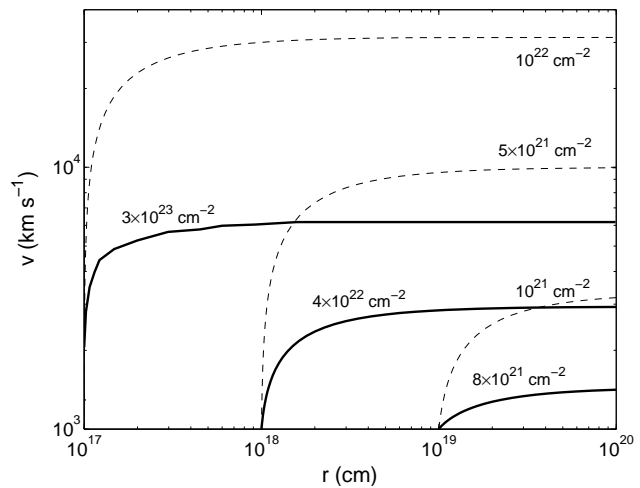} 
\caption{The velocity profile and column densities of model 2 shielded flows for various launching radii. High velocity flows require large column densities and are inconsistent with observations. Flows which occupy only a small fraction of the volume ($\epsilon(r_0)=0.3$) have smaller column densities and higher velocities compared to continuous flows (dashed lines).} 
\label{bshield}
\end{figure}

We have looked for a model which satisfies those conditions and reproduces the typical velocities and column densities of BAL flows. We assumed an initial velocity of $1000~{\rmn {km~s}^{-1}}$ and $U_x(r_0)=1$ (the overall properties are insensitive to these numbers) and calculated velocity profiles for several launching radii $r_0$. The results are shown in figure \ref{bshield}). The column density of the flow is obtained by integrating over the density profile.  Our calculations show that in order to reach high velocities, the flow has to be launched close to the central source ($r_0<10^{17}~{\rmn {cm}}$) which results in a large column density ($\gg 10^{22}~{\rmn {cm}^{-2}}$). Heavier shielding and smaller $U_x(r_0)$ yield similar results.

The above spherically symmetric continuous flow models cannot account for the observed dynamics and spectral features of BAL flows since they result in column densities larger than observed. One possible solution suggested by Arav \& Li (1994) is that such flows consist of small clouds that occupy only a small fraction, $\epsilon$, of the entire volume. These authors assume a continuity equation of the form
\begin{equation}
\epsilon \rho r^2v={\rmn {const}},~\rho \propto r^{-2}.
\label{econt}
\end{equation}
Self-consistent velocity solutions for such flows that are launched
from $\leq 10^{18}~{\rmn {cm}}$ with $\epsilon(r_0)=0.3$ are shown in
figure \ref{bshield} (dashed lines). Such models are consistent with
observations in terms of their column density and terminal
velocity. While alternative solutions to the apparent discrepancy between
models and observations will not be considered here,  we note that
 it is entirely possible that different parts of the flow suffer
different shielding.  The full treatment of such multi-phase flows is beyond the
scope of this paper.

\subsection{Line locking}

Next we study the effect of line-locking on the flow dynamics and spectral features for the geometry specified by equation \ref{econt} and for a small filling factor $\epsilon(r)$.

\begin{figure}
\includegraphics[width=84mm]{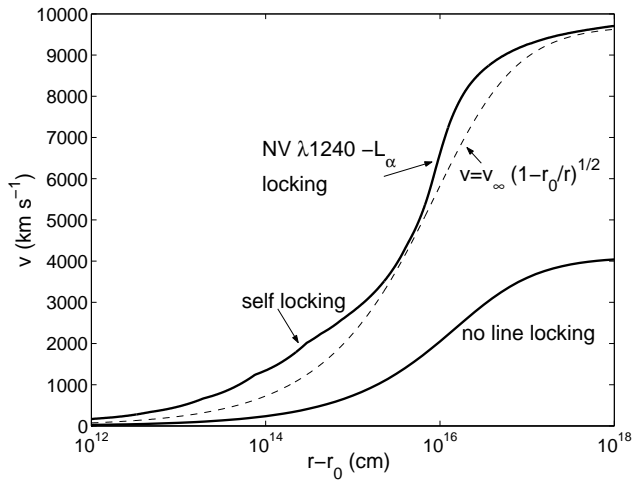}
\caption{The velocity profile of a flow exposed to continuum and broad emission lines. A canonical wind solution with the same terminal velocity is shown for comparison (dashed line). Note the large deviations from the canonical model due to self line locking and \nv~$\lambda1240-L_\alpha$ locking. The terminal velocity of the pure continuum accelerated is a factor $\sim 2$ lower compared to the full calculation.} 
\label{vghost}
\end{figure}

Figure \ref{vghost} compares the velocity profiles of model 3 shielded flows which are exposed to a) pure continuum and b) continuum and emission lines. The initial parameters are: $U_x=0.5,~r_0=3\times 10^{16}~{\rmn {cm}},~\epsilon(r_0)=0.03,~v_0=30~{\rmn {km~s}^{-1}}$. The first and most important result is that the terminal velocity when emission lines are included is a few times higher than the terminal velocity when the flow is exposed only to the AGN continuum. The effect is more pronounced for low velocity, optically thin flows (see CN03). 

\begin{figure}
\includegraphics[width=84mm]{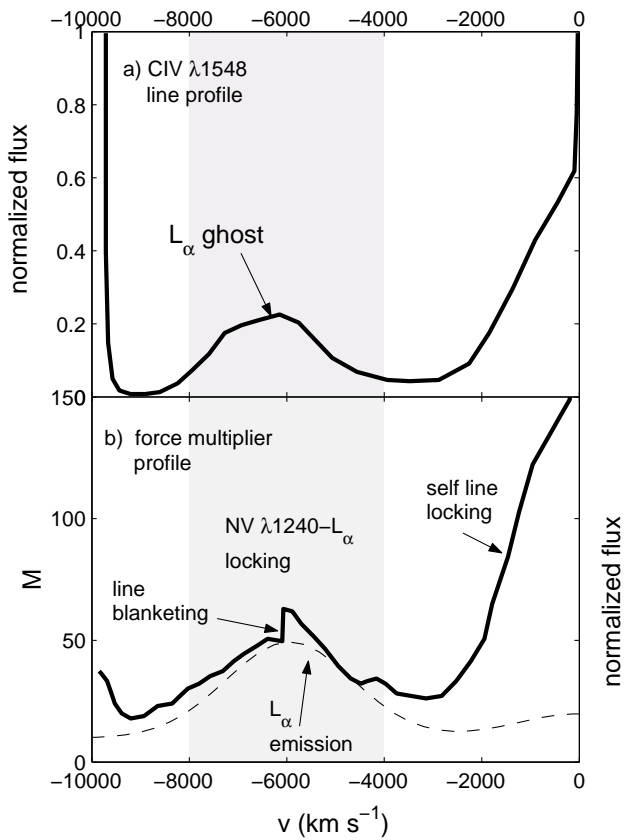} 
\caption{The formation of the ``ghost of $L_\alpha$'' feature. (a) the
  \civ~$\lambda 1549$ line profile showing the double trough feature
  (Arav 1996) (b) The radiation pressure force as a function of the
  flow velocity with the effects of self line locking and \nv~$\lambda
  1240-L_\alpha$ locking. Note that the radiation pressure force
  traces the $L_\alpha$ emission line profile (shown in arbitrary
  units as dashed line) and the \civ~$\lambda 1549$ trough. The sharp
  drop in $M$ at $v=-6000~{\rmn {km~s}^{-1}}$ is due to \nv~$\lambda 1240-L_\alpha$ blanketing. The shaded area highlights the velocity range where \nv~$\lambda 1240-L_\alpha$ locking is dynamically important.} 
\label{lghost}
\end{figure}

The dynamical effect of the \nv~$\lambda 1240-L_\alpha$ locking is
very important. The velocity profile for a flow which is affected by
line locking is shown in figure \ref{vghost} where it is compared to a
canonical wind velocity profile $v=v_\infty(1-r_0/r)^{1/2}$ (e.g.,
Castor, Abbott, \& Klein 1975; hereafter CAK) with the same terminal
velocity but without line locking (dashed line; this model requires
different initial conditions). The resulting velocity profile is
considerably different: At low velocities ($\ltorder 1000~{\rmn
  {km~s}^{-1}}$), self line locking dominates and the acceleration is
larger than the CAK solution. At intermediate velocities ($1000
\ltorder v< \ltorder 4000~{\rmn {km~s}^{-1}}$) the exact calculation
accelerates more slowly than the CAK model since the latter must be
more optically thin in order to reach the same terminal velocity. At
larger velocities ($\gtorder 4000~{\rmn {km~s}^{-1}}$), \nv~$\lambda
1240-L_\alpha$ locking becomes important and the flow accelerates
faster than the CAK solution. The absorption line profile of
\civ~$\lambda 1549$ which results from this velocity profile is shown
in figure \ref{lghost}a. This line profile traces the distribution of
the radiation pressure force (figure \ref{lghost}b). At low velocities
its optical depth per unit velocity, $\tau_v({\mbox{\civ}}~\lambda1549)$, is
small due to the rapid acceleration by self-line locking. As the
acceleration decreases, $\tau_v({\mbox{\civ}}~\lambda1549)$ increases. For
$v\gtorder 3000~{\rmn {km~s}^{-1}}$, the \nv~$\lambda 1240$ line is exposed to the enhanced flux level of the $L_\alpha$ emission line, the radiation pressure force rises, and the \civ~$\lambda 1549$ optical depth decreases (roughly tracing the $L_\alpha$ emission profile). Once $v\gtorder 6000~{\rmn {km~s}^{-1}}$, \nv~$\lambda 1240-L_\alpha$ blanketing becomes important, $M$ declines and the line profile saturates. 

The calculated line profile is not unique to \civ~$\lambda 1549$ and appears also in other lines. Such line profiles have been observed in several BALQSOs and were termed  ``the ghost of $L_\alpha$'' (Arav 1996). Our calculations show that this feature is very sensitive to the abundance of nitrogen  and to the UV flux above the \hi~ edge. In particular, the feature is not created for optically thin \hi~ edges. (see also Arav 1996).

\section{Conclusions and Summary}

We have presented detailed photoionization and dynamical calculations for shielded flows in AGN. This allows us to answer some of the questions raised in section 1:
\begin{enumerate}
\item
  The peculiar shape of the shielded ionizing continuum introduces profound changes in the ionization structure of all flows. Most notable are the lowering of the level of ionization and the almost uniform ionization structure of most metals in highly ionized flows. This may elevate the need for multi-zone models to explain the wide range of ionization levels observed in BAL flows (Turnshek et al. 1996).
\item
The flow dynamics is determined by its level of ionization and the shape of the ionizing continuum. It is therefore highly sensitive to the properties of the shield and to the UV flux above the Lyman edge. Moderate shielding is most effective in driving the gas to high velocities. Heavy shielding, which absorbs a large fraction of the UV and X-ray flux, results in lower velocities compared to unshielded flows. The calculations show that self line locking and \nv~$\lambda 1240-L_\alpha$ line locking can have a substantial effect on the flow dynamics and spectral appearance. Internal radiation pressure can considerably affect the dynamics of optically thick, subsonic flows.  
\item
Spherically expanding flows cannot account, self-consistently, for the typically observed velocities and column densities of BAL flows. One possible explanation explored here is a small filling factor flow that occupies only a small fraction of the volume. This helps to explain the formation of BAL troughs that extend over several decades in velocity.
\item
Our calculations  show that it is possible to explain observed BAL features using detailed non-Sobolev modelling.  Specifically, we were able to account for  the shape of the ``$L_\alpha$ ghost''  observed in some BALQSOs. This opens a new avenue for the determination of flow properties such as the location, geometry, ionization structure, and mass flow rate.
\end{enumerate}

We acknowledge financial support by the Israel
Science Foundation grant no. 545/00, and the Jack Adler Chair of Extragalactic
Astronomy.

\end{document}